\documentclass[aps,twocolumn,epsfig,showpacs]{revtex4}
\usepackage{graphicx}
\usepackage{epsfig}
\begin{document}
\title{Force induced stretched state: Effects of temperature}
\author{Sanjay Kumar and Garima Mishra}
\affiliation{Department of Physics, Banaras Hindu University,
     Varanasi 221 005, India \\}
\begin{abstract}
A model of self avoiding walks with suitable constraint has been developed 
to study the effect of temperature on a single stranded DNA 
($ssDNA$) in the constant force ensemble. Our exact calculations for small
chains show that the extension (reaction co-ordinate)  may increase or decrease 
with the temperature depending upon the applied force. The simple model developed here
which incorporates semi-microscopic details of base direction 
provide an explanation of the force induced transitions in
$ssDNA$ as observed in experiments.
\end{abstract}
\maketitle

In recent years, single molecule force spectroscopy ($SMFS$) has made it possible
to observe force induced transitions in single molecules \cite{sinmol,cof,ubm,bloom,busta}. 
These experiments
also provide information about, for example, transcription and replication processes
involving DNA, mechanical and elastic properties of biopolymers, functional and structural 
properties of proteins etc. Moreover, these experiments also present a platform 
to verify theoretical predictions based on  models 
developed in the framework of statistical mechanics \cite{bhat99,nelson,maren2k2,kbs,kg}.
\begin{figure}[t]
\includegraphics[height=1.5in,width=1.5in]{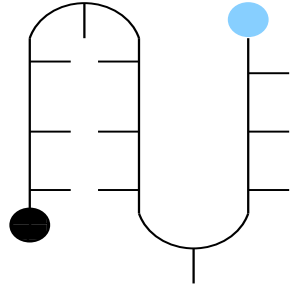}
\caption{(color on line) The schematic representation of a ssDNA where bases are on the 
links of the strand with short stubs representing 
the direction of the hydrogen bonds. 
This figure also shows that once two 
bases are bonded, no further hydrogen bonds can be formed with these bases.
The small black circle indicates that one end of the $ssDNA$ is kept fixed while
a force $F$ may be applied at the other end.}
\label{fig-1}
\end{figure}
In many biological processes there is a large conformational change and the temperature
plays a crucial role. Therefore, efforts of $SMFS$ experiments have now been shifted to study the effect of
the temperature on these processes keeping the force constant \cite{danilow,mao,danilow1}. 
In a recent paper, Danilowicz {\it et al.} \cite{danilow1} studied the elastic properties of the single
stranded DNA ($ssDNA$) and showed that the temperature has significant impact on the force
extension curve. In the low force regime, they found that the extension 
increases with the temperature.  By changing the solvent condition they could show that
the increase in the extension is due to the disruption of hairpins. In fact using 
the Poland-Scheraga ($PS$) 
model of double-stranded DNA ($dsDNA$) \cite{ps} and the modified freely jointed chain model 
(mFJC) \cite{zhang,montanari,dessinges} of a polymer, 
they could nicely represent the force-extension
curve in the low force regime. However, for the higher forces, none of these models 
could explain
the outcome of their experiments, where the extension decreases abruptly with the rise 
of the temperature and it appears that there is no clear understanding about it. It was 
suggested that the observed decrease in extension  may be because of the sequence 
dependent secondary structures \cite{danilow1}.

The mechanical properties of biopolymers ( e.g. DNA, proteins), are fairly well understood 
theoretically from applications of the $mFJC$ \cite{zhang,montanari,dessinges} or  worm like chain ($WLC$) models \cite{marko,smith,seol}. Indeed one finds that the force-extension curves obtained 
from these models agree excellently with experiments 
\cite{sinmol,cof,ubm,bloom,busta,danilow,danilow1}. 
As a result, these models have been used as a benchmark to 
compare the outcome of $SMFS$ experiments.  
It is important to point out here 
that the $WLC$ or $FJC$ model ignores the crucial excluded volume effect \cite{degennes}
in its description, and are therefore, not ideal models to probe the entire phase space. 
The $PS$ model \cite{ps} of the DNA does not include the non-native interactions, 
and, so underestimates the entropy of partial bound states and excludes the 
possibility of the formation of hairpins.  Moreover, the $PS$ model does 
not incorporate configurational entropy and hence it is not appropriate in 
the constant force 
ensemble where the  ``stretched state" may be induced by a force \cite{kijg}. 
Thus these models may only give a limited picture of the unzipping and 
stretching transitions.

Here, we adopt (from the single molecule of finite length point of view) a more 
realistic model of ssDNA proposed in Ref. \cite{kgs}. In 
this, a self-attracting self-avoiding walk ($SASAWs$) along with orientation 
of the base pairs has been considered to model the $ssDNA$.  The base pairing takes 
place only when the nucleotides approach each 
other directly without the sugar phosphate strand coming in between as shown in Fig. 1.

This restriction also takes into account that once two bases are bonded, 
no further hydrogen bonds can be formed with these bases.  By applying a force 
at one end of the chain, system undergoes a phase transition from  the zipped or hairpin 
state to the coil 
(extended state). With further rise of the force, the system goes from the extended 
state to the stretched state, a state which is solely induced by the applied force \cite{kijg}.
The force-temperature phase diagram of unzipping
of $dsDNA$ is known in two dimensions  for some simple models \cite{bhat99,maren2k2,kbs}.
However, in three dimensions, (a $ssDNA$ may form hairpin loop) the phase diagram
is yet to be explored.  As of now experiments using $8$ and more bases are available 
\cite{sosg,lfenb,ma} to describe melting and unzipping of DNA
at a coarse grained level. The purpose of this communication is to provide 
exact results of a semi-microscopic model of a short $ssDNA$ and then 
to study the effect of temperature in a constant force ensemble.
In order to see whether the abrupt decrease in extension is a sequence 
dependent effect \cite{danilow1}, we consider two conformational possibilities of a $ssDNA$. 
In the first case, we allow monomers (nucleotides $\bf T$) of half of the chain 
to form base pairs (nucleotides $\bf A$) with the other half of the chain 
(diblock copolymer). Since in most of the experiments
one end of the chain is kept fixed, the ground state conformation resembles the
zipped state of $dsDNA$ as shown in Fig. 2.  However if we  allow interaction 
among (say) the first three 
($\bf T$) and the last three monomers ($\bf A$), we have the possibility 
of formation of hairpins in a $ssDNA$ at low temperature as shown in Fig. 2.
These two models may be considered as two bounds of ssDNA.

\begin{figure}[t]
\includegraphics[width=2.5in]{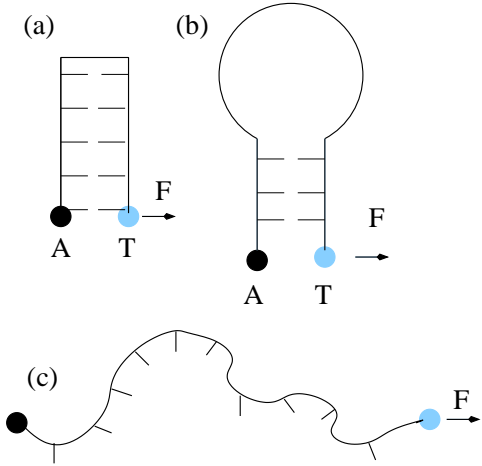} \\
\caption{(color on line) The schematic representation of a $ssDNA$ forming 
(a) the zipped conformation ($dsDNA$). In this case half of the
strand is of $\bf A$ type nucleotides and other half of strands consist
of complementary nucleotides $\bf T$. (b) A hairpin structure of stems of
$3$ bases. Fig. (c) shows that by application of a force  applied to the
one end, the system undergoes from zipped or hairpin state to the extended state.}
\label{fig-2}
\end{figure}
\begin{figure}[t]
\includegraphics[width=3.5in]{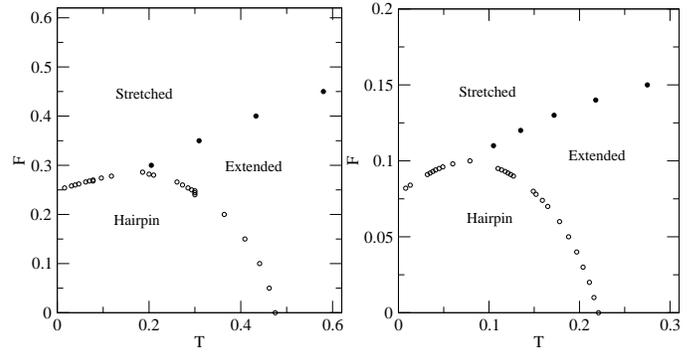} 
\caption{Force-Temperature diagram for (a) zipped case and (b) for DNA hairpin. Solid circles show the crossover regime.
}
\label{fig-3}
\end{figure}
All possible conformations of $SASAWs$ with the base orientations 
have been enumerated on a cubic lattice. Although the time 
involved in enumerating these conformations increases as $\mu^N$ 
where $\mu$ is the connectivity constant of the lattice, using parallel processing we were 
able to enumerate walks of $20$ monomers in $3$ dimensions \cite{exenu}. The thermodynamic 
properties of $ssDNA$ may be determined from the canonical partition sum
\begin{equation}
Z_N = \sum_{p=0}^{n} \sum_{x=0}^{N} C_N(p,x) (e^{-\epsilon/k_B T})^p (e^{-F/k_B T})^x.
\end{equation}
Here $C_N (p, x)$ is the total number of configurations corresponding to a
walk of $N$ steps with $p$ number of base pairs whose end points are at a distance $x$. 
$k_B$, $T$, $\epsilon$ and $F$ are the Boltzmann constant, temperature of the system, 
the base pairing energy and the applied force respectively.  
In the following  calculation we  set $\epsilon/k_B= 1$ and calculate all the 
thermodynamic variables in reduced unit. Here $n$ is the $N/2$ for the zipped case,
while 3 for the hairpin. 
Although no true phase transition can occur in the finite-size single molecule experiments,
 the ``phase transition" observed in such experiments may be considered as 
real if the length of the chain exceeds the characteristic correlation lengths.
We use a sudden change in appropriate average to obtain the different phases in the 
phase diagram \cite{kg,kijg}.  It is pertinent to mention here that thermodynamic limit may be achieved by using the extrapolation technique
developed in \cite{exenu,pks}.  For example, the exact phase diagram of the partial-directed self-avoiding-walks (PDSAWs) is in excellent agreement with the exact enumeration technique \cite{kg,pks}.
It has also been shown that peak values obtained for PDSAWs from fluctuation in non-bonded 
nearest neighbour 
monomers of finite length chain are also in quantitative agreement (within $\pm 0.01$) with exact 
values \cite{kg}. In view of finite-size experiments, we choose this technique so that the complete 
state diagram can be probed exactly.  The force-temperature  diagram
for $ssDNA$ for the zipped and the hairpin cases are shown in Fig. 3. It
is evident from these plots that unzipping force decreases with the rise 
of temperature. Moreover, the existence of re-entrance at low temperature
and the stretched state at a high force observed here are already 
discussed in other context (unzipping and unfolding) and its physical 
origin is known \cite{maren2k2,maren,kijg}. The upper line shown here 
(crossover regime) has been obtained in the constant force ensemble 
which is absent in constant temperature ensemble \cite{maren}.

Here, we focus our studies on the behavior
of the force-extension curve at low temperatures as well as in the high force limit, where
 Monte Carlo and models discussed above fail.
The average extension may be obtained from the relation 
\begin{equation}
<x>=\frac{1}{Z} \sum_{p=0}^{n} \sum_{x=0}^{N} C_N(p,x)x (e^{-\epsilon/k_B T})^p 
(e^{-F/k_B T})^x.
\end{equation}

In Fig. 4, we plot the extension {\it vs} force for the zipped and the hairpin situation at
various temperature respectively. It can be seen from these plots that the extension 
increases with the applied force. This is in agreement with the experiment \cite{danilow1} 
and qualitative understanding is known in terms of dissociation of base
pairs. In constant temperature ensemble, by varying the force one can go from the
zipped state or the hairpin to the extended state. With further rise of the force, 
one finds the stretched state i.e the extension approaches the contour length of 
the $ssDNA$ as seen in the case of stretching of polymers \cite{degennes}. 
However, for a temperature range, we find that these curves cross at a 
critical extension $L_{cross}$. Above this length the force increases with the 
temperature \cite{kijg}. In other words, to keep the extension constant, 
one has to apply more force because the applied force competes with the 
entropy of the chain.

\begin{figure}[t]
\includegraphics[width=3.5in]{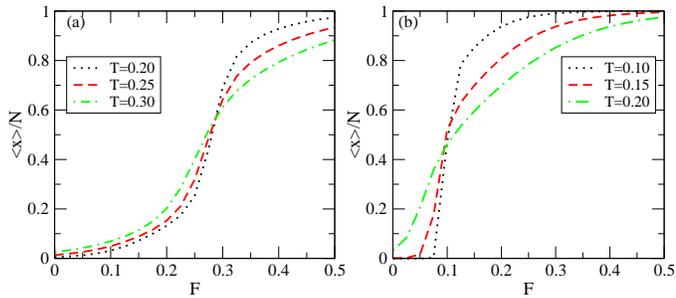} 
\caption{(color on line)The force-extension curve in constant temperature ensemble (a) for
zipped state and (b) for DNA hairpin. One can see the extension approaching the contour
length when the force is varied.
}
\label{fig-4}
\end{figure}

It is now known that different ensembles may give different results
in single molecule experiments \cite{busta}. 
In order to see the effect of the temperature in 
 a constant force ensemble, we plot the extension {\it vs} 
temperature curves in Fig. 5. At low force, the extension increases with the temperature
and the chain acquires the conformation of swollen (extended) state, 
with size $\sim N^\nu$, 
here $\nu$ is the end-to-end distance exponent. In the swollen state (high temperature)
its  value is given by the Flory approximation $\nu \approx 3/(d+2)$ \cite{degennes}.
At high force and low temperature the system attains the stretched state $N^\nu$ with 
$\nu = 1$ and it remains stretched up to a certain temperature. As temperature increases,
the  applied force is not enough to hold the stretched state and the extension falls sharply
to the extended state due to the increased contribution of entropy. This is again in 
agreement with the experiment \cite{danilow1}.  
Since the present model incorporates the configurational 
entropy as well as the formation of hairpin, we are therefore, able to show 
the abrupt decrease in extension with temperature.

\begin{figure}[t]
\includegraphics[width=3.5in]{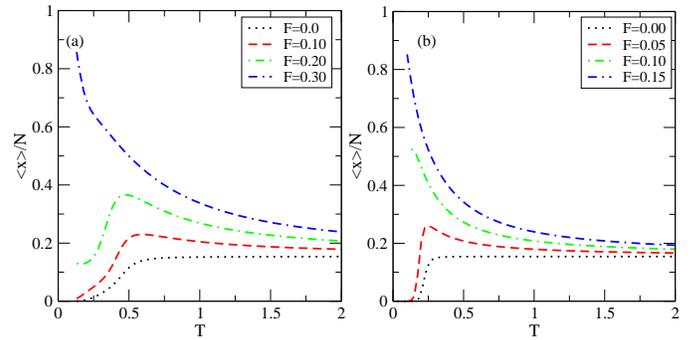} 
\caption{(color on line) Same as of Fig. 4 but in constant force ensemble. 
The abrupt decrease in the extension is obvious from these plots.
}
\label{fig-5}
\end{figure}

If the abrupt fall in extension is the manifestation of entropy, then what is the role
of base-pairing energy? To answer this question, in Fig. 6, we also plot the 
extension {\it vs} temperature curve for a situation where the base-pairing 
energy in the partition function is set equal to $0$. This is identical to 
a non-interacting linear polymer chain in a good solvent. It is surprising to
see that all these curves have similar behavior at high $T$ indicating that
chain is in a swollen state. However, the slope of the 
fall in extension depends on the number of base pairs $p$. It is pertinent to mention here
that at low temperatures, if formation of base-pairing is possible, the system may again 
go to the zipped state as predicted by re-entrance. Since in the experiment, possibility
of forming hairpin is suppressed by the solvent condition, the observed decrease
is due to the entropy.

\begin{figure}[h]
\includegraphics[clip,height=1.8in,width=1.8in]{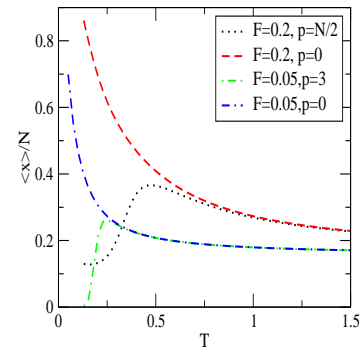}
\caption{(color on line)  Comparison of extension {\it vs} temperature curves with base pairing and without base-pairing. For the zipped state the Number
of base pair is $N/2$ while for hairpin it is $3$. $p=0$ corresponds to the non-interacting
case.
}
\label{fig-6}
\end{figure}
In order to rule out the possibility that the observed effect is due to the small size, 
we revisited, the unfolding of biopolymers, where the data of $55$ steps in $2d$ 
(sufficiently long) allowed us to settle this issue \cite{kijg}.
In Fig. 7a, we plot the average extension with temperature for a non-interacting
polymer (in this case non-bonded nearest neighbor $p = 0$) and interacting polymer 
($p \approx N$). 
This clearly shows that sequence does not play any role as far as decrease
in extension is concerned.  However for interacting polymer, 
the fall is sharper than the non-interacting case. In Fig.7b, we plot the $<x>/N$ with $N$ for different value of $N$.
The collapse of the curves of various lengths of polymer chain at low temperatures indicates 
that the chain is in the ``stretched state" ($\nu =1$) and the observed decrease 
is not a phase transition but a crossover effect.

\begin{figure}[h]
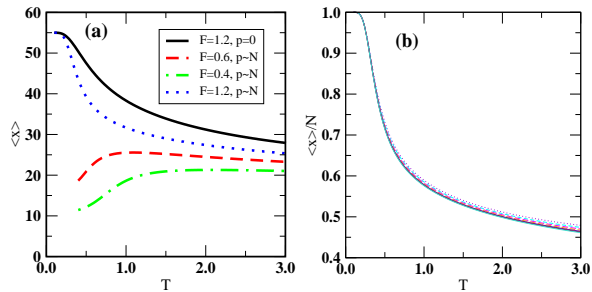

\includegraphics[clip,height=1.5in,width=1.5in]{pre_fig7a.eps}
\includegraphics[clip,height=1.5in,width=1.5in]{pre_fig7b.eps}
\caption{(color on line) (a)  Extension {\it vs} temperature curves in $2d$ for interacting and non-interacting walks of chain length
$55$ in constant force ensemble. 
(b)Extension {\it vs} temperature curves for different length ($N=25,30,35,40,45,50$ 
and $55$) at $F=1.2$. The collapse of data at low temperature
indicates that the chain is in a stretched state.
}
\label{fig-7}
\end{figure}

It would be interesting to see the variation of the entropy with temperature, which can be calculated for the unfolding of biopolymers \cite{kijg} from the following expressions:

\begin{eqnarray}
A &=& - T \ln Z_N (T) \\
S &=& - \left ( \frac{\partial A}{\partial T} \right )
\end{eqnarray}

where $A$ is the Helmholtz free energy \cite{kijg,pks} and we have set the Boltzmann constant $k_B = 1$. In Fig. 8, we show the variation of entropy with temperature for different values of $F$. It is evident from this plot, at high force ($F = 1.2$) and low temperature, the chain is 
in the stretched state. Entropy associated with this state is nearly equal to zero. Rise in temperature brings the system to the high entropic state {\it i. e.} the extended
 state. At zero force, the polymer is in the collapsed state and the entropy associated with this is high but much less than the extended state which is being reflected in Fig. 8. It can also be seen that with increase in temperature system again approach to the extended state. In this figure,
we have also plotted entropy just above and below the phase boundary \cite{kijg} at low temperature. 
A sharp rise in entropy can be seen near the phase boundary around $F =1.0$ 

\begin{figure}[h]
\includegraphics[clip,height=2.in,width=2.in]{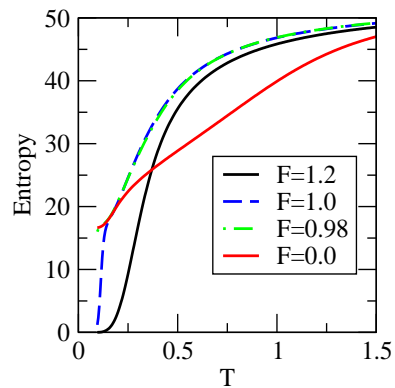}
\caption{(color on line) (a)  Entropy {\it vs} temperature curves in $2d$ for interacting walks of chain length
$55$ steps in constant force ensemble.
At high force $F =1.2$ and  low temperature, the chain is found to
be stretched state and the entropy of the system is nearly equal to zero. As temperature increases
system attains high entropy state. At force equal to zero, one can see the entropy associated with globule is higher than the stretched state but much below than the 
extended state.
The slight variation in force $F=0.98$ and $F=1.0$ around the phase boundary \cite{kijg}
shows sharp fall  in the entropy from the collapsed state to the extended state.
At higher value of $T$, these two curves ($F=1.0$ and $F=0.98$) overlapt to each other.
}
\label{fig-8}
\end{figure}

The conversion of the reduced temperature ($T$) used here and real temperature
($T^*$) measured  in experiments may be obtained through the  relation
$1/T = \epsilon/(kT^*)$ \cite{kg}. It may be noted that present models are a bit different than the $\lambda -$ DNA used in experiments.
In the experiment, it is not known which bases are making hairpins and their
positions. However, in present case we have used two bounds by  putting
either first half and second half complementary to model the zipped case,
or first and last three complementary bases to model DNA hairpin loop.
Since reaction co-ordinates are different and the model is coarse grained,
therefore, a quantitative comparison is not possible. Moreover, effect of salt
concentration, pH of the solvent etc are experimental parameters,
which have been generally ignored in the description of coarse grained
model. Even if, the model developed here with minimum parameters shows the 
qualitative agreement with experiments at low as well as at high force regime
and provide the explanation of the abrupt decrease in the extension.

It would be interesting to see experimentally whether the temperature, 
where abrupt decrease in extension occurs, increases with the force or not. Because the model 
predicts that at high force, the system attains the stretched state and hence close to the upper 
boundary, the force should increase with the temperature. Furthermore, it would be 
useful to repeat these experiments for different chain length at higher force in order
to see the scaling observed in Fig. 7b is a genuine phase transition or a crossover.

In conclusion, we have studied a simple but a realistic model for $ssDNA$ in $3d$ which includes
excluded volume effect, non-native base pairing and the directional nature of the hydrogen
bond. The heterogeneity, intra and inter strand interaction can be incorporated
in the description of this model. It is pertinent to mention here that by replacing 
thymine (T) by uracil (U) and considering intra and inter strand interaction apart from
base pairing, the model may be extended to study the unfolding of RNA \cite{rna}, which is
considered as a step toward the understanding of protein folding. 
For the first time, we have shown that formation of hairpin loop gives rise
the existence of re-entrance in ssDNA. 
In constant force ensemble, the system attains the swollen state (entropy dominated state) 
while in constant temperature ensemble, it acquires the stretched state (force induced state). Our 
results are consistent with the experiment and indicate that the observed 
decrease in extension with temperature may be observed in other $SMFS$ experiments
such as protein unfolding, RNA unfolding and DNA unzipping.

We would like to thank D. Giri for many helpful discussion on the subject. 
Financial supports from UGC and DST, India are acknowledged.

\end{document}